\documentclass[conference]{IEEEtran}
\usepackage{amsmath,amsfonts}
\usepackage{amssymb}
\usepackage{enumerate,letltxmacro}
\usepackage{graphicx}
\usepackage{cite}
\usepackage{subfigure}
\usepackage{multirow}
\usepackage{algorithm}
\usepackage{algorithmic}
\usepackage{mathtools}
\usepackage{hyperref}
\usepackage{verbatim}
\usepackage{xcolor}

\usepackage{array}
\newcolumntype{L}[1]{>{\raggedright\let\newline\\\arraybackslash\hspace{0pt}}m{#1}}
\newcolumntype{C}[1]{>{\centering\let\newline\\\arraybackslash\hspace{0pt}}m{#1}}
\newcolumntype{R}[1]{>{\raggedleft\let\newline\\\arraybackslash\hspace{0pt}}m{#1}}
\hypersetup{draft}

\IEEEoverridecommandlockouts

\newtheorem{theorem}{Theorem}

\newtheorem{definition}{Definition}

\newtheorem{corollary}{Corollary}
\newtheorem{proposition}{Proposition}

\newcommand{\set}[1]{\mathcal{#1}}

\newcommand{\G}{\set{G}}

\title{On the Partition Bound for Undirected Unicast Network Information Capacity}
\author{\IEEEauthorblockN{Mohammad Ishtiyaq Qureshi and Satyajit Thakor}
\IEEEauthorblockN{School of Computing and Electrical Engineering\\
Indian Institute of Technology Mandi, Himachal Pradesh, India}
email: D15063@students.iitmandi.ac.in, satyajit@iitmandi.ac.in
}

\begin{document}
\maketitle
\begin{abstract}
One of the important unsolved problems in information theory is the conjecture that network coding has no rate benefit over routing in undirected unicast networks. Three known bounds on the symmetric rate in undirected unicast information networks are the sparsest cut, the LP bound and the partition bound. In this paper, we present three results on the partition bound. We show that the decision version problem of computing the partition bound is NP-complete. We give complete proofs of optimal routing schemes for two classes of networks that attain the partition bound. Recently, the conjecture was proved for a new class of networks and it was shown that all the network instances for which the conjecture is proved previously are elements of this class. We show the existence of a network for which the partition bound is tight, achievable by routing and is not an element of this new class of networks.
\end{abstract}

\section{Introduction}
Network coding outperforms routing in directed acyclic networks. However, for undirected unicast networks, it was conjectured in 2004 \cite{LiLi04,HarKleLeh04}  that network coding has no rate benefit over routing. This conjecture has found significance in theoretical computer science, see, e.g., \cite{AdlHarETAL06,FarHajLarShi18}. Despite its importance, the conjecture has been verified for only a handful of network instances and families of networks.

Three upper bounds on the symmetric information rate (i.e., the required information rate for each session is the same)
in undirected unicast networks are known: the sparsest cut, the LP bound and the partition bound. The notion of the weighted sparsest cut was given in \cite{Mat85} for graphs, where weights are assigned to edges and can be seen as edge capacity. It was discussed in the context of multi-commodity flow networks in \cite{LeiRao99} and in the context of information networks in \cite{HarKle06}. Sparsest cut is an important upper bound for multi-session undirected unicast commodity capacity but it also holds for information capacity. In \cite{BonBroPatPya12}, it has been shown that the decision version of the unweighted sparsest cut problem is NP-complete where the term unweighted refers to restricting the edge capacity to be unit for every edge. This is shown by considering the problem for a special class of multi-session undirected networks called uniform sparsest cut problem. The LP bound \cite[Chapter 15]{Yeu08}, \cite{HarKle06} has an exponential number of variables as well as an exponential number of constraints and hence it is not computable in practice for even small network instances, see, e.g., \cite[Figure 6]{YinLiLiuWan18}. In \cite{ThaQur18}, we presented a new information-theoretic bound called the partition bound. A parameter is defined as an optimal partition which delivers the partition bound. A recursive formula to compute this parameter and an algorithm are given. The algorithm has exponential complexity. 

This paper presents three results on the partition bound. In Section \ref{sec:Background}, we present the network model and review the undirected unicast network coding conjecture, the partition bound and the associated parameter $\mathrm{opt}(I)$, and known results on the proof of the conjecture for a family of networks. In Section \ref{sec:Main Results}, we present the main contributions of the paper summarized as follows:
\begin{itemize}
\item We show that, as is the case with the unweighted sparsest cut problem, the decision version problem of computing the partition bound is NP-complete.
\item In \cite{AlYon08}, Type-I and Type-II networks were defined and the conjecture was proved for these networks. In \cite{ThaQur18}, these networks were generalized and the conjecture was proved. However, the proof of optimal routing schemes was omitted in both works. We establish a recurrence relation for number of edges in theses networks and give complete proofs of optimal routing schemes.
\item  In \cite{YinLiLiuWan18}, the conjecture was proved for a new class of networks and it was shown that all the network instances for which the conjecture is proved previously are elements of this class. We show the existence of a network for which the partition bound is tight and achievable by routing and is not an element of this class.
\end{itemize}
Finally, a conclusion is presented in Section \ref{sec:Conclusion}.
\section{Background}\label{sec:Background}
An undirected information network is denoted $G=(V,E,I,s,t)$ where $V$ is the set of nodes, $E$ is the set of edges of the form $e=\{u,v\}, u,v\in V$ and $I$ is the set of sessions with $|I|=k$. Assume that each edge has 1 bit (unit) capacity. The source and sink node of a session are described by the mappings $s:I \mapsto V$ and $t:I \mapsto V$ respectively, e.g., session $i$ is available at $s(i)$ (the source node) and demanded at $t(i)$ (the sink node). We consider unicast networks in which a session located at the source node is demanded by exactly one sink node and $s(i)\neq t(i), \forall i\in I$. 

\subsection{The partition bound}

\begin{theorem}[Partition bound, \cite{ThaQur18}]\label{thm:1}
For an undirected network $G=(V,E,I,s,t)$, the symmetric rate of information flow is upper bounded as
 \begin{align}
r &\leq \min_{P} \frac{|E|}{|I|+\sum_{i=1}^{n}|I(P_i,P_i)|}\nonumber\\
&= \frac{|E|}{|I|+\max_{P} \sum_{i=1}^{n}|I(P_i,P_i)|}
\end{align}
where $P$ is a partition of $V$ into independent sets $P_1,\ldots,P_{n}$ and  $I(\alpha,\beta) \triangleq \{i: s(i)\in \alpha, t(i) \in \beta\}, \alpha,\beta \subseteq V$.
\end{theorem}

Let $P^*$ be an optimal partition, i.e.,  a partition such that the total number of source-sink pairs in a same partition set is maximum and opt$(I)$ be a biggest subset of $I$ such that for all $i \in  \mathrm{opt}(I)$ we have $\{s(k),t(k)\}\subseteq P_i$ for some $P_i\in P^*$. Since for some $i\in I$ if $(s(i),t(i))\in E$ then $i$ cannot be in opt$(I)$, it is sufficient to restrict the search for opt$(I)$ in 
\begin{align}
\hat I \triangleq \{i\in I: \{s(i),t(i)\}\not\in E\}.
\end{align}
Now, let the set of neighboring nodes of  $u$ be $\mathrm{ne}(u)=\{v\in V: (v,u)\in E\}$. There are two possibilities for any $k\in \hat I$
\begin{enumerate}
\item $\{s(k),t(k)\}\subseteq P_i$ for some $P_i$ then $\mathrm{opt}(\hat I)=\{k\}\cup \mathrm{opt}(\hat I \setminus\mathrm{conf}(k))$, where conf$(k)$ (abbreviated conflicting subset) is
\begin{align*}
\mathrm{conf}(k)\triangleq& \{l\in \hat I : [t(l)\in \{s(k),t(k)\}, s(l)\in\mathrm{ne}(s(k))] \\
&\text{ or } [s(l)\in\{s(k),t(k)\}, t(l)\in\mathrm{ne}(s(k))]\}
\end{align*}
\item $\{s(k),t(k)\}\not\subseteq P_i$ for any $P_i$ then $\mathrm{opt}(\hat I)=\mathrm{opt}(\hat I\setminus \{k\})$.
\end{enumerate}  
These two possibilities render a recursive formula
\begin{align}
|\mathrm{opt}(I)|&=|\mathrm{opt}(\hat I)|\nonumber\\
&=\max (|\mathrm{opt}(\hat I\setminus \{k\})|, 1+ |\mathrm{opt}(\hat I\setminus\mathrm{conf}(k))|).
\end{align}

Hence, alternatively, the partition bound can be described as
 \begin{align}
r \leq \frac{|E|}{|I|+|\mathrm{opt}(\hat I)|}.
\end{align}

The value $|\mathrm{opt}(I)|$ for a given network can be computed using Algorithm 1 in \cite{ThaQur18}. The complexity of the algorithm is exponential. 
\subsection{Proof of the conjecture for a class of networks}
In this subsection, we review the main results in \cite{YinLiLiuWan18}.
A cut is a partition of $V$, e.g., $\alpha,\alpha^{c}$. The cut-set associated with a partition  $\alpha,\alpha^{c}$ is 
$$\mathrm{cs}(\alpha,\alpha^{c})=\{\{u,v\}\in E: u\in \alpha, v \in \alpha^{c}\}$$ 
There exists a path of length $n$ from node $u$ to $v$ if  there exists nodes $u=v_1,\ldots,v_n=v$ end edges $e_1,\ldots,e_{n-1}$ where $e_i=\{v_i,v_{i+1}\}$. Here, a path from $u$ to $v$, denoted $\mathsf{P}_{u,v}$, is defined as the set of edges involved in it. 
The distance between nodes $u$ and $v$, denoted $d_{u,v}$, is the number of edges in a shortest path connecting the nodes. 
Let $\mathcal P_{u, v}$ be the set of simple paths from node $u$ to $v$.

\begin{definition}
A set of edges $F$ is \textit{orthogonal} to session $i$ if every shortest path from $s(i)$ to $t(i)$ crosses $F$ at most once, i.e., if $|\mathsf{P}_{s(i),t(i)}|=d_{s(i),t(i)}$ then
\begin{align}
|\mathsf{P}_{s(i),t(i)}\cap F|\leq 1.
\end{align}
\end{definition}

\begin{definition}\label{def:compatible}
A set of edges $F$ is called \textit{compatible} with session $i$ if every shortest path from $s(i)$ to $t(i)$ intersects $F$ the minimum number of times among all the $(s(i),t(i))$-paths, i.e., for all $\mathsf{P}_{s(i),t(i)}, \mathsf{P}'_{s(i),t(i)}\in \mathcal P_{s(i) , t(i)}$ such that $|\mathsf{P}_{s(i),t(i)}| = d_{s(i) , t(i)}$ 
\begin{align}
|\mathsf{P}_{s(i),t(i)}\cap F|\leq|\mathsf{P}'_{s(i),t(i)}\cap F|.
\end{align}
\end{definition}

\begin{theorem}[Theorem 3, \cite{YinLiLiuWan18}]\label{thm:YinLiLiuWan18}
Let $\G$ be a family of networks that is closed under edge contractions. If $\G$ contains a network with a coding advantage, there is a network $G\in\G$ satisfying the following properties: 
\begin{enumerate}[\text{ \ }(\textbf{P}1)]
\item every cut-set is not orthogonal to some session; 
\item for any two disjoint cut sets $\mathrm{cs}(\alpha,\alpha^{c})$, $\mathrm{cs}(\beta,\beta^{c})$, if $F = \mathrm{cs}(\alpha,\alpha^{c})\cup \mathrm{cs}(\beta,\beta^{c})$ is compatible with all sessions, then there are at least two source-receiver pairs with a shortest path intersecting $F$ more than twice.
\end{enumerate}
In other words, if $\G$ does not contain a network satisfying (\textbf{P}1) and (\textbf{P}2), network coding is unnecessary in $\G$.
\end{theorem}

Let $\G$ be the class of network as defined in Theorem \ref{thm:YinLiLiuWan18} such that it does not contain a network satisfying (\textbf{P}1) and (\textbf{P}2).
\begin{corollary}[Corollary 3, \cite{YinLiLiuWan18}]\label{cor:YinLiLiuWan18}
Network coding is unnecessary in networks with at most 7 nodes, 3 sessions and uniform link lengths, except one network (see \cite[Figure 6]{YinLiLiuWan18}) for which the network coding capacity is yet unknown. 
\end{corollary}

\section{Main Results}\label{sec:Main Results}
\subsection{On computing $\mathrm{opt}(I)$}
It is well known that the decision version of the sparsest cut problem is NP-complete \cite{BonBroPatPya12}. The computation of linear programming bound \cite{HarKle06} too requires exponential time since it has exponential input constraint size for a given network. In this subsection, we show that the decision version problem of finding $\mathrm{opt}(I)$, \textsc{Optimal-Pairs Decision}, is NP-complete by showing reduction of \textsc{Independent Set} problem which is a well known NP-complete problem.

\begin{quote}
\textsc{Optimal-Pairs Decision} problem:  For $G=(V,E,I,s,t)$, is $\mathrm{opt}(I)$ at least $k$?
\end{quote}

\begin{quote}
\textsc{Independent Set} problem: For $G=(V,E)$, does there exists an independent set of size $k$?
\end{quote}

\begin{theorem}
\textsc{Optimal-Pairs Decision} is NP-complete.
\end{theorem}
\begin{IEEEproof}
First note that \textsc{Optimal-Pairs Decision} is an NP problem: Consider a certificate that is the tuple $(I'\subseteq \mathrm{opt}(I), P)$ where $|I'|=k$ and $P$ is the corresponding partition.
It can be checked in polynomial time whether for each $i\in I'$, $\{s(i),t(i)\}\subseteq P_j$ for some $P_j\in P$.

We prove NP-completeness by reducing an instance of the \textsc{Independent Set} problem to an \textsc{Optimal-Pairs Decision} problem which is known to be NP-complete. The reduction gadget is as follows: For a given graph $G=(V,E)$, fix some node $v \in V$ and let 
$$X\triangleq \{v\} \cup \mathrm{ne}(v).$$ 
Now for each $x\in X$, construct $\bar{G_x}=(V,E,I_x,s,t)$ with $|I_x|=|V|-1$ such that each node in $V\setminus\{x\}$ is a sink for a distinct $i\in I$ and $x$ is the source node for all $i\in I$. It remains to show that there exists an independent set of size $k$ in $G$ if and only if  $\mathrm{opt}(I_x)$ has at least $k-1$ elements in $\bar{G_x}$ for some $x\in X$. 

Assume that there exists an independent set $S$ of size $k$ in $G$. Then observe that there exists an independent set of size $k$ such that it has at least one element of $X$. Hence, assume without loss of generality that there exists some $x \in S\cap X$ (if $S$ does not contain any node from $X$ then we can obtain another independent set of size $k$ from $S$ by removing any one element of it and adding $v$ into it). Now, note that $\mathrm{opt}(I_x)\geq k-1$ since the sessions associated with source-sink pairs such that $x\in S$ is the source and nodes in $S\setminus \{x\}$ are sinks are elements of $\mathrm{opt}(I_x)$ and there are $k-1$ such sessions.

Now assume that there exists $I_x' \subseteq \mathrm{opt}(I_x)$ such that it has $k-1$ elements in $\bar{G_x}$ for some $x\in X$. Given this $I_x'$ with $|I_x'|=k-1$ we can obtain the set 
$$S'_x \triangleq \{y: y=s(i) \text{ or } y=t(i), i\in I'_x\}.$$ 
Note that $S'_x$ is an independent set in $G$ and $|S'_x|=k$.
\end{IEEEproof}

\subsection{Optimal routing schemes for Type-I and Type-II networks}
In this subsection, we first present two expressions, \eqref{eq:rec_rel} and \eqref{eq:x2}, for the cardinality of the edge set for Type-I and Type-II networks and then present proof of routing schemes achieving the partition bound for these networks.
\begin{definition}
\textit{Type-I $n$-partite network} $G=(V,E^{(n)},I,s,t)$ is a complete $n$-partite network with partition sets $P_i, |P_i|\geq 2$ for all $i\in \{1,\ldots,n\}$ such that for every unordered pairs of nodes in a partition set, there is a source-sink pair and there are no other source-sink pairs. 
\end{definition}
\begin{definition}
\textit{Type-II $n$-partite network} $G=(V,E^{(n)},I,s,t)$ is a complete $n$-partite network with partition sets $P_i, |P_i|\geq 2$ for all $i\in \{1,\ldots,n\}$ such that for every unordered pairs of nodes in the network, there is a source-sink pair and there are no other source-sink pairs.
\end{definition}

First, note that for a complete $n$-partite graph with partition sets $P_1,\ldots,P_n$, the cardinality of its edge set $E^{(n)}$ can be expressed via the recurrence relation
\begin{align}
|E^{(n)}|=|E^{(n-1)}|+|P_{n}|\sum_{i=1}^{n-1}|P_i|\label{eq:rec_rel}
\end{align}
and hence, 
\begin{align}
|E^{(n)}|=\sum_{i=2}^{n}\left(|P_i|\sum_{j=1}^{i-1}|P_j|\right).\label{eq:x1}
\end{align}
Alternatively, note that for any two partitions $P_i$, $P_j$ there are $|P_i||P_j|$ number of edges between them and hence,
\begin{align}
|E^{(n)}|=\sum_{\{i,j\}\subseteq\{1,\ldots,n\}: i\neq j}|P_i||P_j|.\label{eq:x2}
\end{align}
We remark that there is an error in the expression of the cardinality of the edge set of Type-I and Type-II networks in \cite{ThaQur18}.
\begin{theorem}\label{thm:Type-I}
For Type-I $n$-partite networks, the partition bound
 is attainable by a routing scheme and hence the conjecture holds for all Type-I $n$-partite networks.
\end{theorem}
\begin{IEEEproof}
The partition bound for a Type-I $n$-partite network is
\begin{align}
\frac{|E^{(n)}|}{2\sum_{i=1}^n |I(P_i,P_i)|}.
\end{align}
We prove the statement by induction. 

\textit{\underline{Base case ($n=2$):}} Consider a bipartite Type-I network with partition sets $P_1$ and $P_2$. 
Assume that $|P_i|\geq 2$. The partition bound is
\begin{align}
&\frac{|E^{(2)}|}{2(|I(P_1,P_1)|+|I(P_2,P_2)|)}\\
=&\frac{|E^{(2)}|}{|P_1|(|P_1|-1)+|P_2|(|P_2|-1)}
\end{align}
Assume that each edge has the capacity
\begin{align}
|P_1|(|P_1|-1)+|P_2|(|P_2|-1).
\end{align}
Now, we show a routing scheme achieving the symmetric rate $|E^{(2)}|$. First, note that each source-sink pair in $P_1$ has $|P_2|$ disjoint 2-hops paths via nodes in $P_2$. For each source-sink pair in $P_1$, transmit 1 bit through each 2-hops path, i.e., path of length 2. In this manner, each edge is involved in transmitting $|P_1|-1$ bits (since, for a Type-I network, each node in $P_i$ has $|P_i|-1$ total number of sources or sinks). In a similar manner the remaining capacity $|P_2|-1$ of each edge can be used to transmit 1 bit for each source-sink pair in $P_2$ through each 2-hops path via nodes in $P_1$. Thus, we can obtain the routing capacity of $|P_1||P_2|=|E^{(2)}|$ (refer to \eqref{eq:x1} or \eqref{eq:x2}).

\textit{\underline{Induction step:}} Assume that the statement is true for $n$-partite Type-I networks. Now consider a $(n+1)$-partite Type-I network with partition sets $P_1,\ldots,P_{n+1}$. Then in its $n$-partite Type-I subnetwork with the partition sets $P_1,\ldots,P_n$, the rate 
\begin{align}
\frac{|E^{(n)}|}{\sum_{i=1}^{n}|P_i|(|P_i|-1)}
\end{align}
is achievable by routing. 
Now, assume that each edge in the $(n+1)$-partite Type-I network has the capacity
\begin{align}
C_e={\sum_{i=1}^{n+1}|P_i|(|P_i|-1)}.
\end{align}
There are $|P_{n+1}|$ disjoint 2-hops paths vie the nodes in $P_{n+1}$ for each source-sink pair in $P_j$. All source-sink pairs in each $P_j, j\in\{1,\ldots,n\}$ have attained (in the $n$-partitie network by induction hypothesis) the symmetric rate of 
\begin{align}
&\frac{C_e|E^{(n)}|}{\sum_{i=1}^{n}|P_i|(|P_i|-1)}=\frac{C_e |E^{(n)}|}{C_e-(|P_{n+1}|(|P_{n+1}|-1))}
\end{align}
But towards proving the statement we need to show that each source-sink pair in the $n+1$-partite network attains the symmetric rate $|E^{(n+1)}|$ bits in total. Hence, for each $P_j, j\in\{1,\ldots,n\}$, the remaining bits need to be sent for each source-sink in $P_j$ via the 2-hops paths through nodes in $P_{n+1}$ are 
\begin{align*}
&|E^{(n+1)}|- \frac{C_e |E^{(n)}|}{C_e-(|P_{n+1}|(|P_{n+1}|-1))}\\
=&\frac{|E^{(n+1)}|[C_e-|P_{n+1}|(|P_{n+1}|-1)]- C_e |E^{(n)}|}{[C_e-(|P_{n+1}|(|P_{n+1}|-1))]}\\
=&\frac{C_e[|E^{(n+1)}|-|E^{(n)}|]-|E^{(n+1)}||P_{n+1}|(|P_{n+1}|-1)}{[C_e-(|P_{n+1}|(|P_{n+1}|-1))]}\\
=&\frac{C_e[|P_{n+1}|\sum_{i=1}^{n}|P_i|]-|E^{(n+1)}||P_{n+1}|(|P_{n+1}|-1)}{[C_e-(|P_{n+1}|(|P_{n+1}|-1))]}\\
=& \Delta
\end{align*}
where we have used the recurrence relation for $|E^{(n+1)}|$, see \eqref{eq:rec_rel}. Assume that $\Delta$ bits are sent in this manner:
$\Delta/|P_{n+1}|$ bits via each 2-hops path through nodes in $P_{n+1}$. In a Type-I network, each node in $P_j$ has $|P_j|-1$ total number of sources or sinks and thus, each edge carries $(|P_j|-1)\Delta/|P_{n+1}|$ bits. The remaining capacity of each edge between nodes in $P_j$ and $P_{n+1}$ is
\begin{align}
&C_e-\frac{(|P_j|-1)\Delta}{|P_{n+1}|}.
\end{align}
Hence, 
\begin{align}
\frac{[C_e|P_{n+1}|-(|P_j|-1)\Delta]}{|P_{n+1}|}\times \frac{|P_j|}{|P_{n+1}|-1}
\end{align}
bits can be sent for each source-sink pair in $P_{n+1}$ via the 2-hops paths through the nodes in $P_{j}$. Thus, the total number of bits which can be sent for each source-sink pair in $P_{n+1}$ via the 2-hops paths though the nodes in the partition sets $P_1,\ldots,P_n$ is
\begin{align}
\sum_{j=1}^n\frac{[C_e|P_{n+1}|-(|P_j|-1)\Delta]|P_j|}{|P_{n+1}|(|P_{n+1}|-1)}.\label{eq:x3}
\end{align}
It remains to show that the above expression \eqref{eq:x3} is indeed $|E^{(n+1)}|$.

\begin{align*}
&\sum_{j=1}^n\frac{[C_e|P_{n+1}|-(|P_j|-1)\Delta]|P_j|}{|P_{n+1}|(|P_{n+1}|-1)}\\
=& \frac{C_e|P_{n+1}| \sum_{i=1}^n |P_i| - \Delta \sum_{i=1}^n |P_i|(|P_i|-1)}{|P_{n+1}|(|P_{n+1}|-1)}\\
=& \frac{C_e|P_{n+1}| \sum_{i=1}^n |P_i| - \Delta [C_e-(|P_{n+1}|(|P_{n+1}|-1))]}{|P_{n+1}|(|P_{n+1}|-1)}\\
=& \frac{C_e|P_{n+1}| \sum_{i=1}^n |P_i| - C_e[|P_{n+1}|\sum_{i=1}^{n}|P_i|]}{|P_{n+1}|(|P_{n+1}|-1)}+|E^{(n+1)}|\\
=&0+|E^{(n+1)}|
\end{align*}
Thus, the routing scheme attains the partition bound and in doing so it fully utilizes the capacity of each edge.
\end{IEEEproof}

\begin{theorem}\label{thm:Type-II}
For Type-II $n$-partite networks, the partition bound
 is attainable by a routing scheme and hence the conjecture holds for all Type-II $n$-partite networks.
\end{theorem}
\begin{IEEEproof}
The partition bound for a Type-I $n$-partite network is
\begin{align}
&\frac{|E^{(n)}|}{\sum_{\{i,j\}\subseteq\{1,\ldots,n\}: i\neq j}|I(P_i,P_j)|+2\sum_{i=1}^n |I(P_i,P_i)|}\\
=&\frac{|E^{(n)}|}{\sum_{\{i,j\}\subseteq\{1,\ldots,n\}: i\neq j}|P_i||P_j|+{\sum_{i=1}^{n}|P_i|(|P_i|-1)}}\\
=&\frac{|E^{(n)}|}{|E^{(n)}|+{\sum_{i=1}^{n}|P_i|(|P_i|-1)}}.
\end{align}

Now, assume that each edge in the $n$-partite Type-I network has the capacity
\begin{align}
C_e={|E^{(n)}|+\sum_{i=1}^{n}|P_i|(|P_i|-1)}.
\end{align}

By Theorem \ref{thm:Type-I}, we can attain the rate $E^{(n)}$ for each session $i$ such that $\{s(i),t(i)\}\subseteq P_j$ for some $P_j$ and this will utilize $\sum_{i=1}^{n}|P_i|(|P_i|-1)$ bits capacity for each edge in the network. Hence, each edge can carry further $|E^{(n)}|$ bits. This fraction of the capacity can be used to transmit  $|E^{(n)}|$ bits for each session $i$ such that $s(i)\in P_j, t(i)\in P_l, j\neq l$; source and sink node for each of such sessions are directly connected by an edge. Thus the partition bound is achieve by considering the routing scheme for Type-I networks described in Theorem \ref{thm:Type-I} and then superpositioning the flow $|E^{(n)}|$ for each session such that the respective source-sink pair is one hop away.
\end{IEEEproof}

\subsection{ A result on sets of networks: Type-I and $\G$}

In this section, we show the existence of a Type-I $n$-partite network that satisfies the properties (\textbf{P}1) and (\textbf{P}2) in Theorem \ref{thm:YinLiLiuWan18}. Hence, there exists a network that is not in $\G$ and for which the partition bound is tight and achievable by a routing scheme.

The following proposition shows that not all  Type-I $n$-partite networks where $n\geq3$ satisfy both (\textbf{P}1) and (\textbf{P}2).
\begin{proposition}\label{prop:1}
There exist Type-I $n$-partite networks, $n\geq3$, for which (\textbf{P}1) is violated.
\end{proposition}
\begin{IEEEproof}
The smallest example is the following: Consider Type-I $3$-partite network with $P_i=\{s(i),t(i)\}, i\in\{1,2,3\}$ and hence $|P_i|=2, i\in\{1,2,3\}$. Consider the cut-set $\mathrm{cs}(\alpha,\alpha^c)$ where $\alpha=\{s(i), i=1,2,3\}$. Then it is straightforward to verify that $\mathrm{cs}(\alpha,\alpha^c)$ is orthogonal to all sessions and hence the property (\textbf{P}1) is violated. 

In fact, using the similar cut-set construction, it can be shown that all Type-I $n$-partite networks with $P_i=\{s(i),t(i)\}, i\in\{1,\ldots,n\}$ and $|P_i|=2, i\in\{1,\ldots,n\}$ violate (\textbf{P}1).
\end{IEEEproof}

\begin{theorem}
There exists a Type-I $n$-partite network, $n\geq 3$, that satisfies (\textbf{P}1) and (\textbf{P}2).
\end{theorem}
\begin{IEEEproof}
Consider the Type-I $3$-partite network in Figure \ref{fig:3-pNet}. It has 7 nodes $v_1,\ldots,v_7$, 16 edges and 5 sessions, i.e., $I=\{1,\ldots,5\}$. The mappings $s$ and $t$ are depicted in the figure, e.g., $s(1)=v_2$ and $t(1)=v_1$. Note that, in a type-I network, every shortest path from $s(i)$ to $t(i)$ is of length 2 for each session $i$. 

\begin{figure}[htbp]
\centering
  \includegraphics[scale=.65]{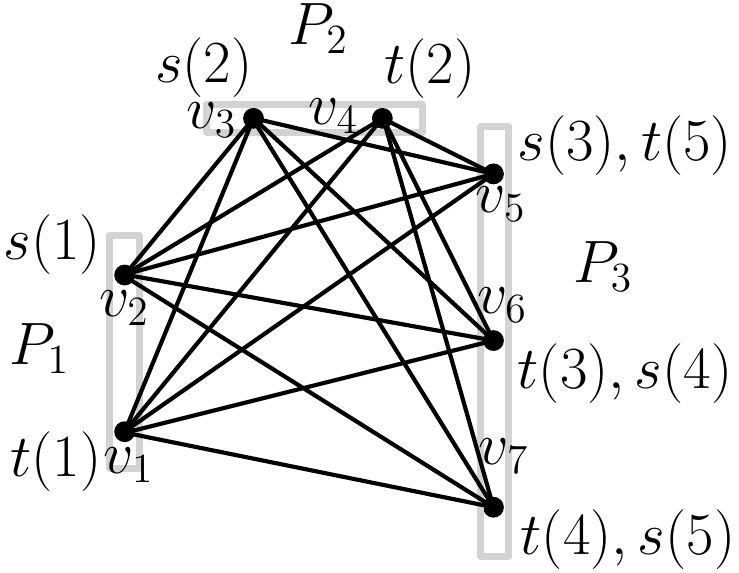}
  \caption{A Type-I 3-partite network.}\label{fig:3-pNet}
\end{figure}

Note that there are $2^{7-1}-1=63$ possible distinct cut-sets of the form $\mathrm{cs}(\alpha,\alpha^c)$ in the network (similar bi-partitions of a set are also considered in \cite{ThaChaGra17} but in a different context). Without loss of generality, assume that $\alpha$ contains the node $v_1$. Thus we need to consider all cut-sets $\mathrm{cs}(\alpha,\alpha^c)$ such that $\alpha$ contains the node $v_1$ and it is a proper subset of $V$ (since $\alpha,\alpha^c$ form a partition, they cannot be the empty set by definition). Also, there are inherent symmetries in the network: permutations on the elements of the sets $\{v_1,v_2\}$, $\{v_3,v_4\}$, $\{v_5,v_6,v_7\}$, $\{\{v_1,v_2\},\{v_3,v_4\}\}$ results in essentially the same network since these permutations only changes the mappings $s$ and $t$ but do not affect the network topology.

In columns 1-3 of Table \ref{tab:1}, we list the sets $\alpha$ for all such cut-sets $\mathrm{cs}(\alpha,\alpha^c)$, corresponding non-orthogonal sessions, and symmetric cases of $\alpha$. We note that for each cut-set there exists at least one non-orthogonal session and hence (\textbf{P}1) is satisfied. 

Now, for each possible subset $\alpha$ in column 1, column 4 shows all possible subsets $\beta$ of $V$ such that $\mathrm{cs}(\alpha,\alpha^c)\cap \mathrm{cs}(\beta,\beta^c)=\emptyset$, i.e., the cut-sets are disjoint. For each tuple $(\alpha,\beta)$ such that $\mathrm{cs}(\alpha,\alpha^c)\cap \mathrm{cs}(\beta,\beta^c)=\emptyset$, we verify that $\mathrm{cs}(\alpha,\alpha^c)\cup \mathrm{cs}(\beta,\beta^c)$ is not compatible with all sessions; details are omitted (refer to Definition \ref{def:compatible} and note that every shortest path from a source to the respective sink is of length 2 and $F_1\triangleq \mathrm{cs}(\alpha,\alpha^c)\cup \mathrm{cs}(\beta,\beta^c)$ covers some shortest paths but not all shortest paths from $s(i)$ to $t(i)$ for at least one $i$. This means that $F_1$ is not compatible with all sessions. For example, consider $\alpha=\{v_1\}$ and $\beta=\{v_2\}$. Then, $\mathrm{cs}(\alpha,\alpha^c)\cup \mathrm{cs}(\beta,\beta^c)$ is the set of all edges between the nodes in $P_1$ and $P_1^c$ and for session $3$, a shortest path via $P_1$ has non-empty intersection with $F_1$ whereas a shortest path via $P_2$ has non-empty intersection with $F_1$. Thus, $F_1$ is not compatible with session 3). Hence, (\textbf{P}2) is vacuously true for the Type-I $3$-partite network in Figure \ref{fig:3-pNet}.
\begin{table}[]
\caption{Summary of results on cut-sets for the network in Figure \ref{fig:3-pNet}}
\label{tab:1}
\begin{tabular}{|L{1.6cm} | L{.65cm} | L{3.85cm} |  L{.9cm} |}
\hline
$\alpha$          & Non-orth. $i\in I$ & Symmetric cases of $\alpha$                                                                                                            & $\beta$                     \\ \hline
$\{v_1\}$           & $2$, $3$, $4$, $5$               & -                                                                                                                                      & $\{v_2\}$                     \\ \hline
$\{v_1,v_2\}$         & $1$, $2$, $3$, $4$, $5$             & -                                                                                                                                      & -                           \\ \hline
$\{v_1,v_3\}$         & $3$, $4$, $5$                 & $\{v_1,v_4\}$                                                                                                                              & -                           \\ \hline
$\{v_1,v_5\}$         & $2,4$                   & $\{v_1,v_6\}$, $\{v_1,v_7\}$                                                                                                                   & -                           \\ \hline
$\{v_1,v_2,v_3\}$       & $1$, $3$, $4$, $5$               & $\{v_1,v_2,v_4\}$, $\{v_1,v_3,v_4\}$                                                                                                               & -                           \\ \hline
$\{v_1,v_2,v_5\}$       & $1$, $2$, $4$                 & $\{v_1,v_2,v_6\}$, $\{v_1,v_2,v_7\}$                                                                                                               & -                           \\ \hline
$\{v_1,v_3,v_5\}$       & $4$                     & $\{v_1,v_3,v_6\}$, $\{v_1,v_3,v_7\}$, $\{v_1,v_4,v_5\}$, $\{v_1,v_4,v_6\}$, $\{v_1,v_4,v_7\}$                                                                        & -                           \\ \hline
$\{v_1,v_5,v_6\}$       & $2,3$                   & $\{v_1,v_5,v_7\}$, $\{v_1,v_6,v_7\}$                                                                                                               & -                           \\ \hline
$\{v_1,v_2$, $\text{ \ \ \ \ \ }v_3,v_4\}$     & $1$, $2$, $3$, $4$, $5$                 & -                                                                                                                                      & -                           \\ \hline
$\{v_1,v_2$, $\text{ \ \ \ \ \ }v_3,v_5\}$     & $1,4$                   & $\{v_1,v_2,v_3,v_6\},\{v_1,v_2,v_3,v_7\}$, $\{v_1,v_2,v_4,v_5\},\{v_1,v_2,v_4,v_6\}$, $\{v_1,v_2,v_4,v_7\},\{v_1,v_3,v_4,v_5\}$, $\{v_1,v_3,v_4,v_6\},\{v_1,v_3,v_4,v_7\}$                 & -                           \\ \hline
$\{v_1,v_2$, $\text{ \ \ \ \ \ }v_5,v_6\}$     & $1$, $2$, $3$                 & $\{v_1,v_2,v_5,v_7\},\{v_1,v_2,v_6,v_7\}$                                                                                                           & -                           \\ \hline
$\{v_1,v_3,$ $\text{ \ \ \ \ \ }v_5,v_6\}$     & $3$                     & $\{v_1,v_3,v_5,v_7\},\{v_1,v_3,v_6,v_7\}$, $\{v_1,v_4,v_5,v_6\},\{v_1,v_4,v_5,v_7\}$, $\{v_1,v_4,v_6,v_7\}$                                                              & -                           \\ \hline
$\{v_1,v_5,$ $\text{ \ \ \ \ \ }v_6,v_7\}$     & $2$, $3$, $4$, $5$               & -                                                                                                                                      & -                           \\ \hline
$\{v_1,v_2,v_3,$ $\text{ \ \ \ \ \ \ \  }v_4,v_5\}$   & $1$, $2$, $4$                     & $\{v_1,v_2,v_3,v_4,v_6\}$, $\{v_1,v_2,v_3,v_4,v_7\}$                                                                                                       & $\{v_5\}$                     \\ \hline
$\{v_1,v_2,v_3,$ $\text{ \ \ \ \ \ \ \ }v_5,v_6\}$   & $1$, $3$                   & $\{v_1,v_2,v_3,v_5,v_7\}$, $\{v_1,v_2,v_3,v_6,v_7\}$, $\{v_1,v_2,v_4,v_5,v_6\}$, $\{v_1,v_2,v_4,v_5,v_7\}$, $\{v_1,v_2,v_4,v_6,v_7\}$, $\{v_1,v_3,v_4,v_5,v_6\}$, $\{v_1,v_3,v_4,v_5,v_7\}$, $\{v_1,v_3,v_4,v_6,v_7\}$ & -                           \\ \hline
$\{v_1,v_2,v_5,$ $\text{ \ \ \ \ \ \ \ }v_6,v_7\}$   & $1$, $2$, $3$, $4$, $5$             & -                                                                                                                                      & -                           \\ \hline
$\{v_1,v_3,v_5,$ $\text{ \ \ \ \ \ \ \ }v_6,v_7\}$   & $3$, $4$, $5$                 & $\{v_1,v_4,v_5,v_6,v_7\}$                                                                                                                        & -                           \\ \hline
$\{v_1,v_2,v_3,$ $\text{ \ \ \ }v_4,v_5,v_6\}$ & $1$, $2$                   & $\{v_1,v_2,v_3,v_4,v_5,v_7\}$, $\{v_1,v_2,v_3,v_4,v_6,v_7\}$                                                                                                   & $\{v_5\}$, $\{v_6\}$, $\{v_5,v_6\}$ \\ \hline
$\{v_1,v_2,v_3,$ $\text{ \ \ \ }v_5,v_6,v_7\}$ & $1$, $3$, $4$, $5$               & $\{v_1,v_2,v_4,v_5,v_6,v_7\}$, $\{v_1,v_3,v_4,v_5,v_6,v_7\}$                                                                                                   & $\{v_3\}$                     \\ \hline
\end{tabular}
\end{table}
\end{IEEEproof}

Also, note that the network in Figure \ref{fig:3-pNet} does not fall into the family of networks for which the conjecture is proved in Corollary \ref{cor:YinLiLiuWan18}. Thus, we have established the existence of a network not in $\G$ such that the partition bound on the symmetric rate is tight and achievable by a routing scheme and hence the conjecture holds for the network. However, Proposition \ref{prop:1} shows that not all Type-I networks satisfy properties (\textbf{P}1) and (\textbf{P}2).

\section{Conclusion}\label{sec:Conclusion}
We showed that, as is the case with the sparsest cut problem, the decision version of computing the partition bound is NP-complete and gave a complete proof of the optimal routing schemes for Type-I and Type-II networks. Also, We showed that the partition bound is tight and achievable by routing for networks for which the conjecture has not been proved previously. One interesting future direction is to characterize the class of all networks for which the partition bound is tight and achievable by routing.
\section*{Acknowledgment}
This work is supported by SERB, DST, Government of India, under Extra Mural Scheme SB/S3/EECE/265/2016.
\bibliographystyle{ieeetr}
\bibliography{network}

\begin{thebibliography}{10}

\bibitem{LiLi04}
Z.~Li and B.~Li, ``Network coding in undirected networks,'' in {\em Proc. 38th
  Annu. Conf. Inf. Sci. Syst. (CISS)}, 2004.

\bibitem{HarKleLeh04}
N.~J.~A. Harvey, R.~D. Kleinberg, and A.~R. Lehman, ``Comparing network coding
  with multicommodity flow for the k-pairs communication problem,'' Tech. Rep.
  MIT-LCS-TR-964, 2004.

\bibitem{AdlHarETAL06}
M.~Adler, N.~J.~A. Harvey, K.~Jain, R.~Kleinberg, and A.~R. Lehman, ``On the
  capacity of information networks,'' in {\em ACM-SIAM Symp. on Disc. Algo.},
  pp.~241--250, 2006.

\bibitem{FarHajLarShi18}
A.~Farhadi, M.~Hajiaghayi, K.~G. Larsen, and E.~Shi, ``Lower bounds for
  external memory integer sorting via network coding,'' {\em CoRR},
  vol.~abs/1811.01313, 2018.

\bibitem{Mat85}
D.~W. Matula, {\em Concurrent Flow and Concurrent Connectivity on Graphs},
  p.~543–559.
\newblock USA: John Wiley $\&$ Sons, Inc., 1985.

\bibitem{LeiRao99}
T.~Leighton and S.~Rao, ``Multicommodity max-flow min-cut theorems and their
  use in designing approximation algorithms,'' {\em J. ACM}, vol.~46,
  pp.~787--832, Nov. 1999.

\bibitem{HarKle06}
N.~Harvey, R.~Kleinberg, and A.~Lehman, ``On the capacity of information
  networks,'' {\em IEEE Trans. Inform. Theory}, vol.~52, pp.~2345--2364, Jun.
  2006.

\bibitem{BonBroPatPya12}
P.~Bonsma, H.~Broersma, V.~Patel, and A.~Pyatkin, ``The complexity of finding
  uniform sparsest cuts in various graph classes,'' {\em Journal of Disc.
  Algo.}, vol.~14, pp.~136 -- 149, 2012.

\bibitem{Yeu08}
R.~W. Yeung, {\em Information Theory and Network Coding}.
\newblock Springer, 2008.

\bibitem{YinLiLiuWan18}
X.~{Yin}, Z.~{Li}, Y.~{Liu}, and X.~{Wang}, ``A reduction approach to the
  multiple-unicast conjecture in network coding,'' {\em IEEE Trans. Inform.
  Theory}, vol.~64, pp.~4530--4539, June 2018.

\bibitem{ThaQur18}
S.~{Thakor} and M.~I. {Qureshi}, ``Undirected unicast network capacity: A
  partition bound,'' in {\em IEEE Int. Symp. Inform. Theory}, pp.~196--200,
  July 2019.

\bibitem{AlYon08}
A.~Al-Bashabsheh and A.~Yongacoglu, ``On the {$k$}-pairs problem,'' in {\em
  IEEE Int. Symp. Inform. Theory}, pp.~1828--1832, July 2008.

\bibitem{ThaChaGra17}
S.~{Thakor}, T.~{Chan}, and A.~{Grant}, ``Capacity bounds for networks with
  correlated sources and characterisation of distributions by entropies,'' {\em
  IEEE Trans. Inform. Theory}, vol.~63, pp.~3540--3553, June 2017.

\end{thebibliography}
\end{document}